\documentclass[10pt,a4paper,twocolumn,longbibliography,pra,superscriptaddress]{revtex4-2}

\usepackage{graphicx}
\usepackage{amsmath}
\usepackage{amsfonts,dsfont}
\usepackage{hyperref}
\usepackage{textcomp}

\hypersetup{
colorlinks,
linkcolor=blue,
citecolor=blue,
urlcolor=blue
}

\DeclareMathOperator{\Tr}{Tr}

\def\ket#1{\left| #1 \right\rangle}
\def\bra#1{\left\langle #1 \right|}
\def\ketbra#1#2{\left| #1 \right\rangle\!\left\langle #2 \right |}

\begin{document}
	
\title{Effect of source statistics on utilizing photon entanglement in quantum key distribution}

\author{Radim Ho\v{s}\'{a}k}
\email{hosak@optics.upol.cz}
\author{Ivo Straka}
\affiliation{Department of Optics, Palack\'{y} University, 17. listopadu 12, 77146 Olomouc, Czech Republic}
\author{Ana Predojevi\'c}
\affiliation{Department of Physics, Stockholm University, 10691 Stockholm, Sweden}
\author{Radim Filip}
\author{Miroslav Je\v{z}ek}
\affiliation{Department of Optics, Palack\'{y} University, 17. listopadu 12, 77146 Olomouc, Czech Republic}

\begin{abstract}
A workflow for evaluation of entanglement source quality is proposed. Based on quantum state density matrices obtained from theoretical models and experimental data, we make an estimate of a potential performance of a quantum entanglement source in quantum key distribution protocols. This workflow is showcased for continuously pumped spontaneous parametric down-conversion (SPDC) source, where it highlights the trade-off between entangled pair generation rate and entanglement quality caused by multiphoton nature of the generated quantum states. We employ this characterization technique to show that secure key rate of  down-converted photon pairs is limited to 0.029 bits per detection window due to intrinsic multiphoton contributions. We also report that there exists one optimum gain for continuous-wave down-conversion sources. We find a bound for secure key rate extracted from SPDC sources and make a comparison with perfectly single-pair quantum states, such as those produced by quantum dots.
\end{abstract}

\date{\today}
\maketitle

\section{Introduction}

\subsection{The aim of this paper}

Quantum entanglement enables a multitude of technological leaps in computing and communications. We address the generation of photonic entanglement in a discrete degree of freedom, such as polarization. Realistic entanglement sources often possess a trade-off between entanglement quality and generation rate. Both of these are important in practical applications. For example, optical nonlinear parametric processes such as spontaneous parametric down-conversion (SPDC) can yield high-fidelity polarization entanglement between two photons \cite{Kwiat1999}. However, with increasing generation rate, inherent multi-pair contributions deteriorate the measurable quantum correlations \cite{Takesue2010, Takeoka2015}. On the other hand, sources based on self-assembled quantum dots allow generating polarization entanglement via two energy-degenerate cascades \cite{Huber2014} without systematic multi-pair statistics in the produced states.

Comparing the performance of these entanglement sources is not a clear-cut problem as conventional entanglement measures cannot be applied. The first reason is that quantum states generated by different systems can populate Hilbert spaces of different dimensions. This issue arises for example due to differences in the statistical properties of the corresponding sources. The second reason is that entanglement measures do not reflect entanglement generation rate. That is why we choose to evaluate the entanglement sources by their potential in quantum key distribution (QKD) \cite{Bennett1992,Devetak2005}. Namely, the secure key rate is a quantity that benefits both from measurable strong quantum correlations (a Bell factor) and from high generation rate.

As executing entire QKD protocols for the purpose of entanglement source characterization is quite complex practically, we choose to carry out quantum state tomography instead. The effective density matrices are reconstructed within the informational degrees of freedom, and are affected by both single-pair entanglement imperfections and multi-pair effects. Then, we analyze how well these density matrices would fare in a QKD protocol \cite{Pironio2009}. As we are interested in measuring the potential of a source by itself, we assume that the rest of the QKD protocol does not suffer from further technical limitations or loopholes \cite{Hensen2015Oct,Shalm2015Dec,Giustina2015Dec}. We evaluate the limit of long-key transmission rate $R_\mathrm{key}$ per detection window as a function of coincidence rate $r_\mathrm{C}$.

\subsection{Quantum key distribution}

The field of quantum key distribution exploits quantum correlations to ensure that two distant parties can share messages without their contents being compromised by a possible presence of a malicious party eavesdropping on the communication channel \cite{Bennett1992}. Many QKD protocols for transmission of discretely encoded information rely on random choices of encoding and decoding bases. Perhaps the best known is the protocol BB84 \cite{Bennett1984} whose performance relies on non-classicality of the single-photon source employed \cite{Lasota2017}. The protocol E91 \cite{Ekert1991} incorporated entanglement into QKD, employing random switching between projection bases. Furthermore, if the security is guaranteed even in presence of untrustworthy entanglement source or detection equipment, the protocols can be called device-independent QKD \cite{Acin2007, Vazirani2014, Xu2020, Pirandola2020Dec}.

The major figure of merit in QKD is the minimum average bit rate at which secure key can be transmitted between distant parties \cite{Devetak2005}. Evaluating this rate is generally a task different from assessing the presence of entanglement. Secure key rate itself depends on other factors, for example the quantum bit error rate (QBER) \cite{Pironio2009}, or the capabilities of the eavesdropper. Therefore, it can occur that certain sources of entanglement cannot be used at all for secure QKD through a given channel whereas others might be viable. In order to prove the secure key rate for the sources, the whole protocol would need to be carried out explicitly. Such data is usually not available, because they would require extensive experimental effort. So instead, we make use of quantum tomography data that is commonly available for quantum entanglement sources. This involves reducing the generated states to the informational degrees of freedom by projecting multi-photon contributions into a two-qubit Hilbert space. The resulting two-qubit density matrices are then evaluated in terms of QKD performance. This approach, however, should not be treated as a proof of QKD performance of states containing multiphoton noise, because the multi-photon contribution may have different effect on tomography than in QKD. The secure key rate calculated from the reconstructed two-qubit matrices shows how multi-pair contributions influence the effective purity and entanglement of intrinsically multi-photon states. The secure key rate also offers a way of quantifying the trade-off between generation rate and entanglement quality. It therefore serves to evaluate entanglement of two-qubit density matrices, given their generation rate. Because only the source is being characterized, we consider the rest of the QKD protocol to be ideal, not taking into account aspects such as finite key size or the detection loophole.

\subsection{Entanglement sources}

In this work, we compare two photon-pair sources of polarization entanglement -- continuous-wave spontaneous parametric down conversion (SPDC) \cite{Ling2008} and self-assembled quantum dots from the perspective of entanglement-based QKD. These two physical platforms produce entangled states of different modal structure.

First, we focus on SPDC. It is a nonlinear optical process which produces entangled photon pairs in two optical modes. We assume a continuous-wave pump and the temporal coherence of the photons to be much shorter than the generation rate and detector resolution. We detect the quantum states in the coincidence basis, meaning only simultaneous detections in both modes are recorded. Consequently, the generated signal can be considered a random Poissonian sequence of photon pairs that are entangled in polarization. Such randomness inevitably leads to detecting multiple pairs within one detection window. This becomes more prominent with increasing detection window and with a higher gain of the source.

We elected CW pumping in favor of pulsed SPDC, because the effect of the multi-pair contributions is much lower at the same generation rate. Our analysis indicates that if there is one pump pulse in every detection window, the multi-pair contributions of both CW and pulsed regimes have the same effect on secure key rate. However, the pulse repetition frequency is usually not that high. The detection window width is only limited by the temporal jitter of the detectors, which can easily be $< 10^{-9}$~s even for non-cryogenic detectors. On the other hand, the typical repetition frequency of pulsed SPDC pumps is in the order $\sim 10^{8}~\mathrm{s}^{-1}$. This means that CW gets an order-of-magnitude advantage in secure key rate. This holds even in the light of the most recent advances in cryogenic detector resolution (3~ps) \cite{Wang2019} and SPDC pump frequency (43~GHz) \cite{Zeiger2019Jun}.

Multi-photon nature is inherent to SPDC and it has been studied in the context of single-photon sources \cite{Straka2014, Predojevic2014, Somaschi2016}, in quantum information processing \cite{OBrien2007, Varnava2008, Jennewein2011} and quantum key distribution \cite{Ma2007, Holloway2013} protocols. SPDC for QKD has also been investigated with respect to noise and its effect on the detection efficiency required to achieve provable protocol security \cite{Ho2020}.

The second physical platform involves quantum dots. They act as semiconductor embedded quantum emitters and allow the optical generation of photon pairs via decay of a biexciton. The energy degeneracy of two biexciton cascades leads to a superposition of two decay paths and thus to entanglement of the emitted photon pairs. Excitation and de-excitation of such cascades is a Rabi cycle that is pumped by a $\pi$-pulse \cite{Jayakumar2013}. Therefore, a quantum dot produces no more than one entangled photon pair at a time \cite{Predojevic2014} and with near-unity generation efficiency. This is the key difference between quantum dots and SPDC. However, it is much more challenging to reach a good collection efficiency of the photons, which means that entangled pairs are usually extracted from quantum dot sources at low effective rates.

We provide a model for SPDC entanglement sources pumped by a continuous wave (CW) laser. Then, we compare the model with experimental SPDC data and current state-of-the-art quantum dot sources. We find that the secure key rate using CW SPDC is fundamentally bounded, whereas quantum dot sources are capable of surpassing this bound.

\section{Secure key rate}
For our investigation we are assuming an entanglement-based QKD protocol, the security of which was analyzed in Refs.~\cite{Acin2007,Pironio2009}. Polarization-encoded photonic qubits will be assumed. The protocol relies on Alice and Bob sharing a two-qubit entangled state. Alice can choose one of three measurements $\mathrm{A}_0$, $\mathrm{A}_1$, $\mathrm{A}_2$ to perform on her qubit, and Bob can choose from two measurements $\mathrm{B}_1$ and $\mathrm{B}_2$ to perform on his qubit.
The measurement results $a_i, b_j$ are binary: $+1$~or~$-1$. Furthermore, they fulfill the following condition:
\begin{equation}
\label{eq:uniform_marginals}
\langle a_i \rangle = \langle b_j \rangle = 0 \quad \forall i,j.
\end{equation}
The results of measurements $\mathrm{A}_0$ and $\mathrm{B}_1$ are used to extract the raw key, whereas the measurements $\mathrm{A}_1$, $\mathrm{A}_2$, $\mathrm{B}_1$, and $\mathrm{B}_2$ are used to calculate the CHSH polynomial $S$ \cite{Clauser1969Oct}. The protocol is only secure for $S$ that violates the classical inequality, e.g. $2 < S \le 2\sqrt2$.
In general, the rate $r$ of the secure key in a given QKD protocol is very difficult to ascertain \cite{Camalet2020, Tan2019, Kaur2020} as this asks for a very specific definition of the security level attained. Instead, we limit ourselves here to an estimate of the lower bound on secure key rate per photon pair. This is given by a quantity called the Devetak-Winter rate $r_\mathrm{DW}$ \cite{Devetak2005, Pironio2009}
\begin{equation}
\label{eq:rdw}
r \ge r_\mathrm{DW} = I(\mathrm{A}_0:\mathrm{B}_1) - \chi(\mathrm{B}_1:\mathrm{E}),
\end{equation}
where $I(\mathrm{A}_0:\mathrm{B}_1)$ is the mutual information between Alice and Bob, and $\chi(\mathrm{B}_1:\mathrm{E})$ is the Holevo quantity \cite{Holevo1973} between Bob and Eve. For the studied protocol, the mutual information can be expressed as
\begin{equation}
\label{eq:pironio_mutual_inf}
I(\mathrm{A}_0:\mathrm{B}_1) = 1 - h(Q),
\end{equation}
while the Holevo quantity is bounded as follows:
\begin{equation}
\label{eq:pironio_holevo}
\chi(\mathrm{B}_1:\mathrm{E}) \le h \left(\frac{1 + \sqrt{\left(S/2\right)^2 - 1}}{2}\right).
\end{equation}
Here $h$ denotes the binary entropy function $h(Q) = -Q\log_2{Q} - (1-Q) \log_2{(1-Q)}$, $S$ the CHSH polynomial, and $Q$ the quantum bit error rate (QBER) defined as the probability of opposite measurement results when the bases $\mathrm{A}_0$ and $\mathrm{B}_1$ are used
\begin{equation}
\label{eq:qber}
Q = P(a\neq b | \mathrm{A}_0, \mathrm{B}_1).
\end{equation}
Substituting (\ref{eq:pironio_mutual_inf}) and (\ref{eq:pironio_holevo}) into (\ref{eq:rdw}) leads to the following bound:
\begin{equation}
\label{eq:pironio_rate}
r_\mathrm{DW} = 1 - h(Q) - h\left(\frac{1 + \sqrt{\left(S/2\right)^2 - 1}}{2}\right).
\end{equation}

This quantity represents the minimum ratio of the bits used as the secure key relative to the number of entangled pairs that were detected. To introduce secure key rate, we need to define the coincidence rate $r_\mathrm{C}$ as the rate of detected photon pairs per detection window. For a measurement with data acquisition subdivided into $N_\mathrm{win}$ windows, where a total of $N_\mathrm{C}$ coincidences were registered, the coincidence rate is
\begin{equation}
\label{eq:CW_coinc_rate}
r_\mathrm{C} = \frac{N_\mathrm{C}}{N_\mathrm{win}}.
\end{equation}
Then, the secure key rate becomes
\begin{equation}\label{eq:Rkey}
R_{\mathrm{key}} = r_\mathrm{DW} \cdot r_\mathrm{C},
\end{equation}
quantifying the minimum number of secure key bits transferred per one detection window that serves as the basic unit of time.

Because $r_\mathrm{DW}$ is a function of $S$ and $Q$, let us calculate these quantities provided that we have the effective quantum state $\rho$. The correlation tensor $T_\rho$ and the positive symmetric tensor $U_\rho$ \cite{Horodecki2009} are first calculated,
\begin{eqnarray}
\label{eq:corr_tensor_elements}
T_{\rho,ij} &=& \mathrm{Tr}\left[\rho \cdot \left(\sigma_i \otimes \sigma_j\right)\right],\quad i,j=1,2,3,\\
U_\rho &=& T_\rho^\mathrm{T} T_\rho,
\end{eqnarray}
where $\sigma_i$ are the Pauli matrices. Consequently the three eigenvalues of $U_\rho$ are sorted in a descending order, $\lambda_1 \geq \lambda_2 \geq \lambda_3$. The best possible values of $S$ \cite{Horodecki1995} and $Q$ are then
\begin{align}
\label{eq:bmax}
S_\mathrm{max} &= 2\sqrt{\lambda_1 + \lambda_2},\quad \\
Q_\mathrm{min} &= \frac{1 - \sqrt{\lambda_1}}{2}.\quad \text{(see Appendix \ref{sec:cookbook})}
\end{align}
Appendix~\ref{sec:cookbook} shows the corresponding optimal configuration of bases $A_{0,1,2}$, $B_{1,2}$, and provides some additional information including an experimental guide to setting the waveplates.

\section{SPDC entanglement source}
SPDC sources at very low gains produce maximally entangled two-qubit states with a multi-photon-pair component which is negligible in the context of the QKD protocol. However, at higher interaction gains, the multi-pair contribution emerges and starts to deteriorate the quality of entanglement \cite{Takeoka2015,Klimov2014,Mueller2016}. This leads to diminishing $r_\mathrm{DW}$. On the other hand, increasing gain leads to higher brightness and the rate $r_\mathrm{C}$ increases.

To study the effect of SPDC multi-photon component on potential performance in QKD, we used a continuously pumped entanglement source with a variable coincidence window. Although the SPDC gain should ideally be controlled by pump power, in the CW case it can be equivalently controlled by the coincidence window $\tau$ used for data acquisition. We leveraged this to study a broader range of scenarios with more ease. With longer coincidence windows, there is a chance of photons generated as products of independent SPDC processes to contribute to the coincidence count. As the individual processes are independent and very fast, the amount of pairs collected by the detectors within the coincidence window of length $\tau$ obeys the Poissonian statistics with a mean pair number $\bar{n}$ proportional to $\tau$.

The model described in Appendix~\ref{sec:CW_tomography_model} enables us to see how the detection of multiple independent copies of the state $\rho_0$ affects the reconstructed two-qubit density matrix $\rho$ effectively describing the state. The whole model is parameterized by the mean photon pair number $\bar{n}$ per detection window and overall optical transmittances in Alice's and Bob's part of the physical setup $\eta_\mathrm{A}$ and $\eta_\mathrm{B}$, respectively. The transmittances consist of signal collection efficiency, transmission loss, and detection efficiency. As we are interested in characterizing entanglement sources exclusively, we do not consider additional noise and loss present in the optical communication channel.

It is possible to choose the state $\rho_0$ from an experimentally obtained reconstruction
of a real quantum entangled state produced by low-gain SPDC. This approach allows to account for realistic experimental imperfections present in the generated quantum states.
To allow for analytical insight, we consider $\rho_0$ to be one of the Bell states, which we will denote $\rho_\mathcal{B}$. Then, the maximum-likelihood estimate of $\rho$ is a mixture of the Bell state and white noise,
\begin{equation}
\label{eq:model_rho}
\tilde{\rho}_\mathcal{B} = \left(1-\kappa\right) \rho_\mathcal{B} + \kappa \frac{1}{4} \mathds{1}\otimes\mathds{1},
\end{equation}
where $\mathds{1}$ is a unity matrix. The parameter $\kappa$ depends on the physical parameters $\bar{n}$, $\eta_\mathrm{A}$, and $\eta_\mathrm{B}$ (see Appendix~\ref{sec:CW_tomography_model}).

For a density matrix $\tilde\rho_\mathcal{B}$ of the form (\ref{eq:model_rho}), both $S$ and $Q$ are related to $\kappa$ as follows:
\begin{equation}
\label{eq:S_and_Q_kappa}
S = 2 \sqrt{2\left(1 - \kappa\right)^2},\quad Q = \frac{\kappa}{2}.
\end{equation}
From these, $r_\mathrm{DW}$ can be calculated using \eqref{eq:pironio_rate}.

\begin{figure}[t]
	\centering\includegraphics[width=\columnwidth]{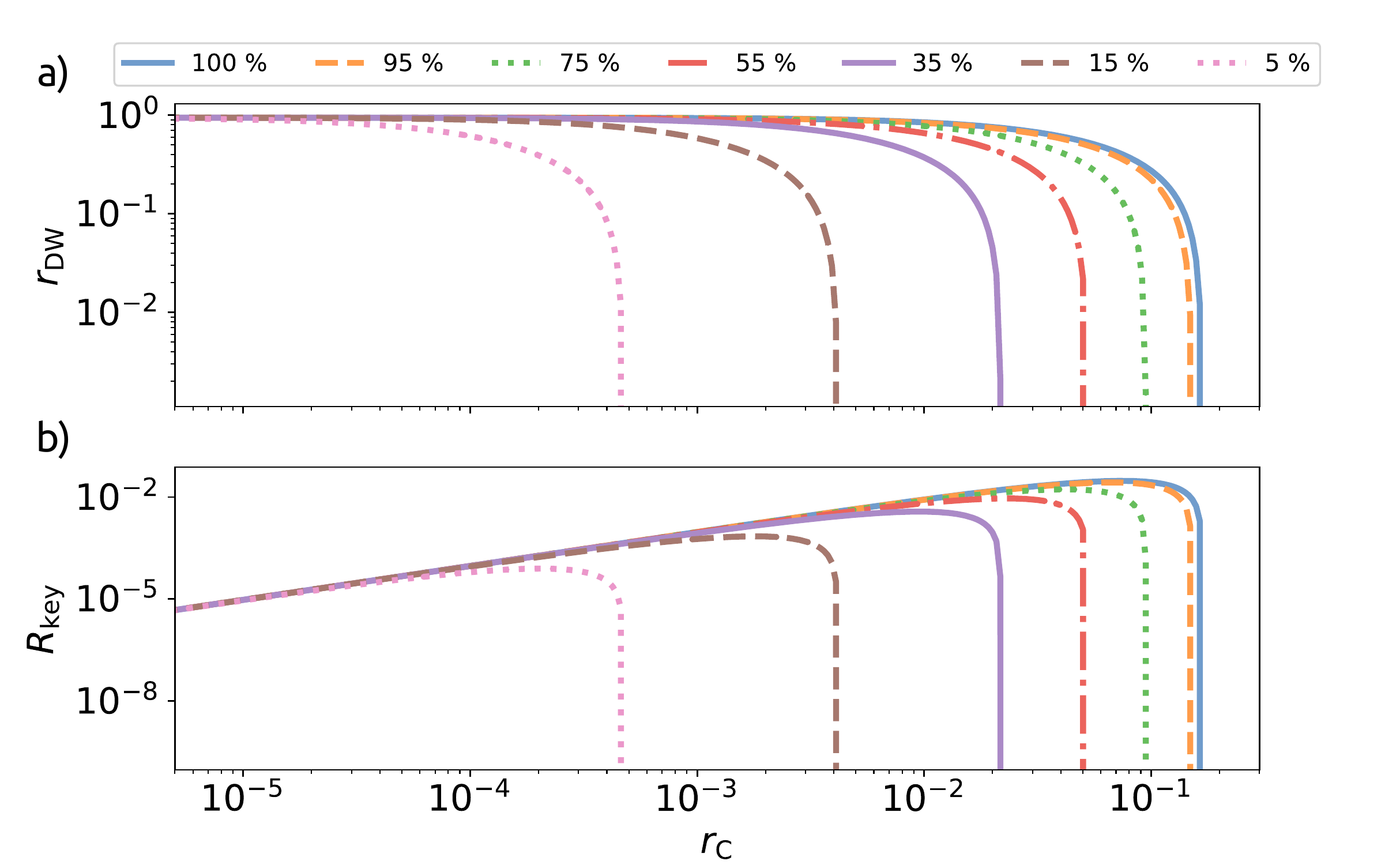}
	\caption{a) the lower bound on secure key rate $r_\mathrm{DW}$ as a function of coincidence rate $r_\mathrm{C}$ for a low-gain CW SPDC entanglement source. b) the key rate $R_\mathrm{key}$ as a function of $r_\mathrm{C}$. Calculations for various values of symmetric transmittance $\eta_\mathrm{A} = \eta_\mathrm{B} = \eta$ are shown as differently colored lines. The loss-free case $\eta = 100\%$ is shown as the rightmost solid blue line, and represents a fundamental limitation of SPDC entanglement source performance in the QKD protocol. The quantities $r_\mathrm{DW}$, $R_\mathrm{key}$, and $r_\mathrm{C}$ are a function of $\bar{n}$, as given by \eqref{eq:pironio_rate}, \eqref{eq:Rkey}, \eqref{eq:S_and_Q_kappa}, \eqref{eq:kappa_approx}, and \eqref{eq:rC_approx}; for exact formulae see Appendix~\ref{sec:CW_tomography_model}.}
	\label{fig:CW_skr_vs_rate_with_losses}
\end{figure}

Our model addresses the trade-off between entanglement quality, reflected by $r_\mathrm{DW}$, and entangled pair quantity, which corresponds to $r_\mathrm{C}$. This is shown in Fig.~\ref{fig:CW_skr_vs_rate_with_losses}a. At low $r_\mathrm{C}$, $r_\mathrm{DW}$ maintains a very high value close to one. As $r_\mathrm{C}$ increases, however, $r_\mathrm{DW}$ starts to deteriorate quickly, until the QKD protocol ceases to be secure. The underlying mechanism behind this gradual degradation of QKD security lies in the multi-photon component. Multiple pairs inside one window result chiefly in three-photon and four-photon events. As the one-qubit reductions of \eqref{eq:model_rho} are always maximally mixed, the extra multi-photon contribution corresponds to white noise. The quantities $r_\mathrm{C}$ and $r_\mathrm{DW}$ are multiplied to obtain $R_\mathrm{key}$ which is the main figure of merit. When plotted against $r_\mathrm{C}$, as shown in Fig.~\ref{fig:CW_skr_vs_rate_with_losses}b, a linear increase in key rate can be seen at first, until a peak is reached for a certain $r_\mathrm{C}$, after which key rate drops quickly. In a two-dimensional space of quality ($R_\mathrm{key}$) and quantity ($r_\mathrm{C}$) axes, the zero-loss case $\eta=1$ bounds the area that is accessible to continuous SPDC sources.

The model can be analyzed in another way. When we expand the exact formula for $\kappa$ (see Appendix \ref{sec:CW_tomography_model}) into a Taylor series, we can see that in the low-gain regime $\bar{n} \ll 1$, 
\begin{align}\label{eq:kappa_approx}
\kappa &\approx \frac{\bar{n}}{1+\bar{n}},	\\\label{eq:rC_approx}
r_\mathrm{C} &\approx \bar{n} \cdot \eta_\mathrm{A} \eta_\mathrm{B}.
\end{align}
This means that the quantum state $\rho$ depends very little on transmittances. Moreover, using these relations, we can  see that $R_\mathrm{key}$ is a factorizable function: $R_\mathrm{key} \approx \eta_\mathrm{A}\eta_\mathrm{B} \bar{n}r_\mathrm{DW}(\bar{n})$. This allows us to optimize the key rate with respect to $\bar{n}$,
\begin{equation}\label{eq:Rkeyopt}
R_\mathrm{key}^\mathrm{opt} \approx 0.029 \cdot \eta_\mathrm{A} \eta_\mathrm{B} \quad \text{for} \quad \bar{n}_\mathrm{opt} \approx 0.0737.
\end{equation}
With this particular value of $\bar{n}$, the corresponding key rate will always be within 0.2 \% of the real maximum value for the given transmittances.

\begin{figure}[t]
	\centering\includegraphics[width=\columnwidth]{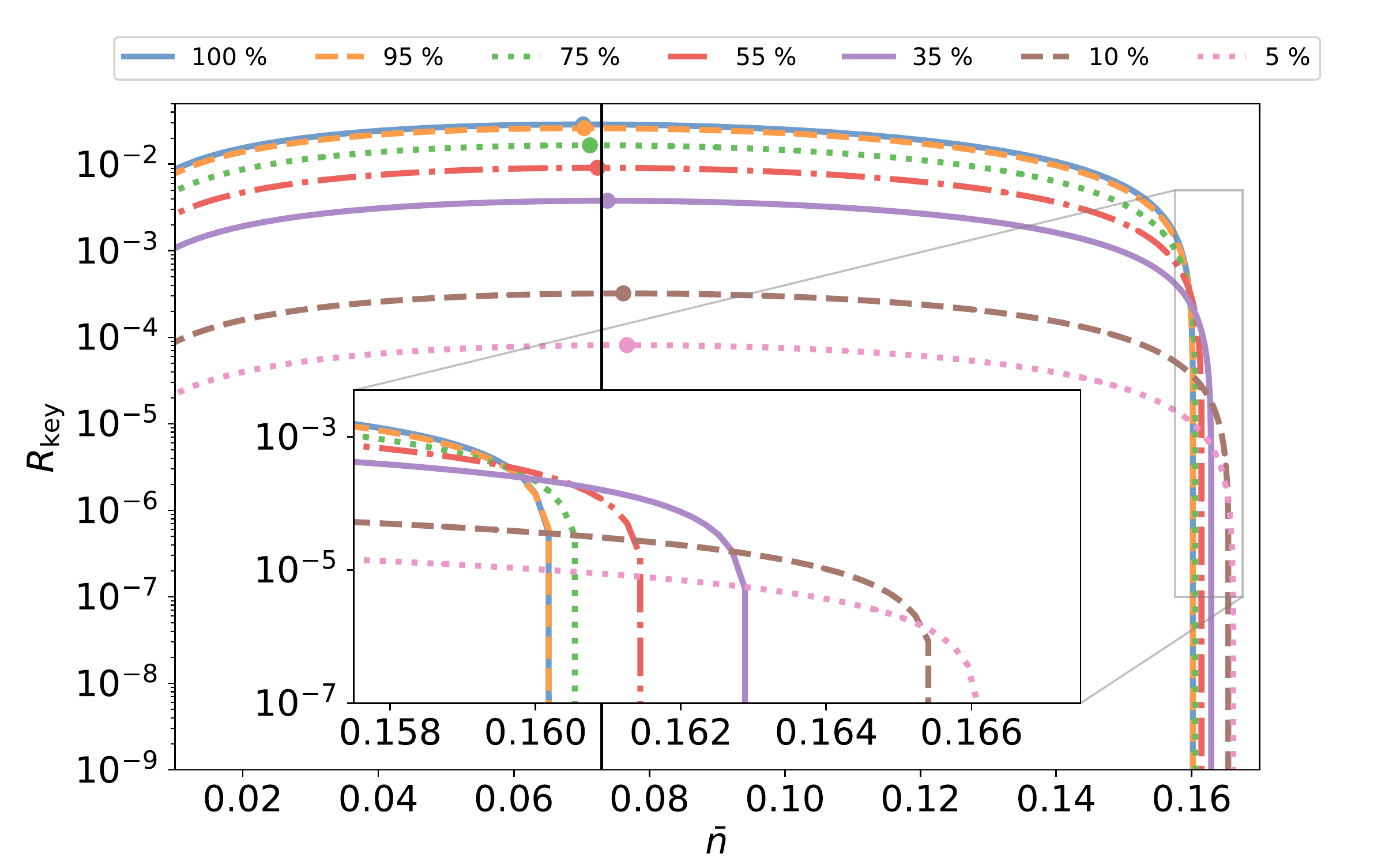}
	\caption{Secure key rate $R_\mathrm{key}$ as a function of mean photon pair number $\bar n$ for various amounts of symmetric transmittance $\eta_\mathrm{A} = \eta_\mathrm{B} = \eta$. The optimal values of key rate for each transmittance are shown as dots. The black vertical line represents the $\bar{n} = 0.0737$ for which the $R_\mathrm{key}$ is within 0.2 \% of the maximum $R_\mathrm{key}$ for the given transmittance. The inset shows how key rate starts to diminish as $\bar n$ approaches the critical value.}
	\label{fig:key_rate_vs_meanN}
\end{figure}

Fig.~\ref{fig:key_rate_vs_meanN} shows that the transmittances are primarily a scaling factor for $R_\mathrm{key}$ and highlights the optimal points \eqref{eq:Rkeyopt}. Fig.~\ref{fig:key_rate_vs_meanN} also shows that around $\bar{n} = 0.16$, the key rate starts dropping sharply. The maximal value of $\bar{n}$ giving a non-zero key rate is 0.166839, in the limit of zero transmittance.

\begin{figure}[t]
	\centering\includegraphics[width=\columnwidth]{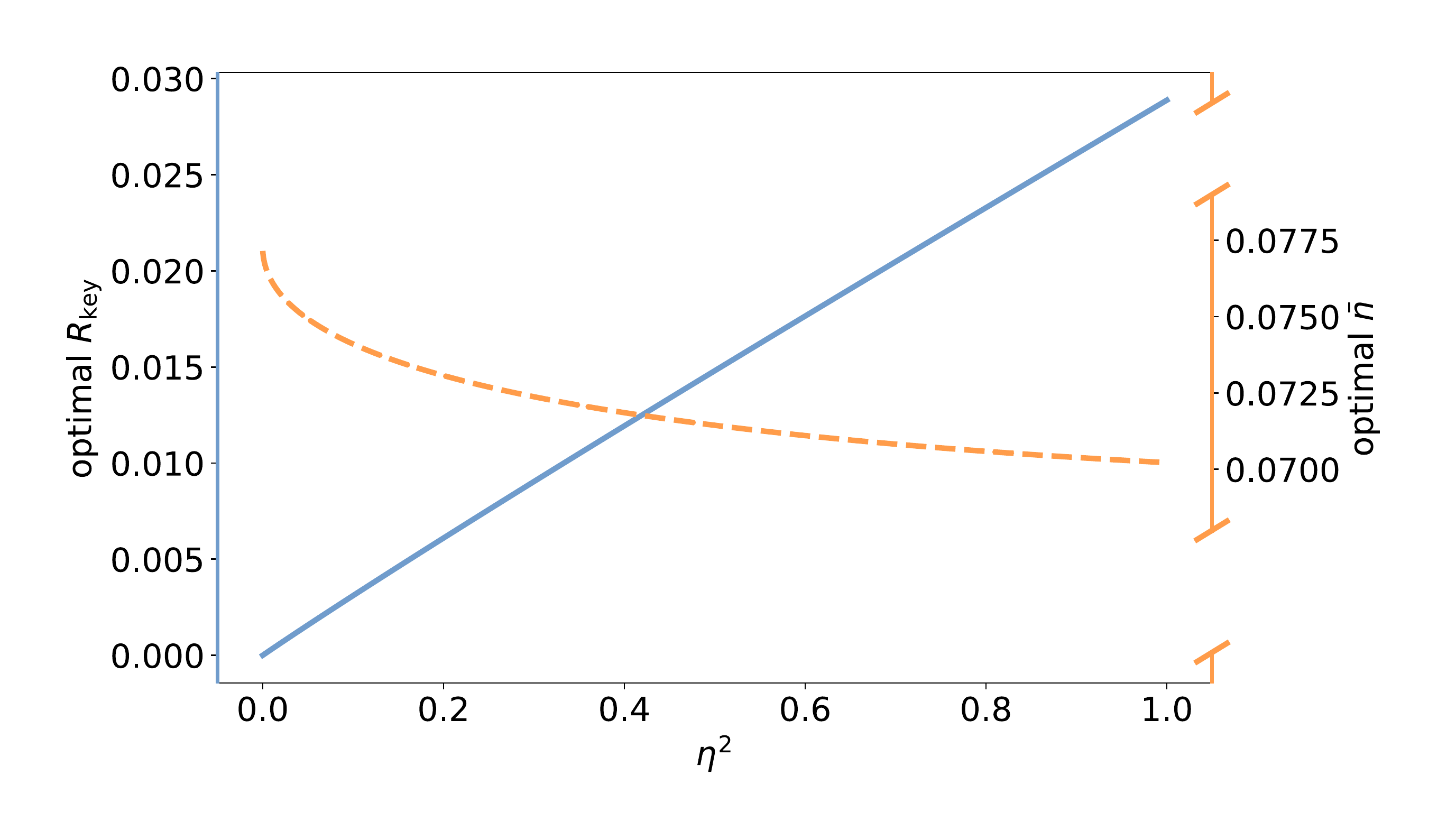}
	\caption{The optimal achievable key rate $R_\mathrm{key}$ (blue line) and the corresponding mean photon pair number $\bar{n}$ (orange dashed line) for a given two-mode transmittance $\eta^2$. Symmetric single-mode transmittance $\eta_\mathrm{A} = \eta_\mathrm{B} = \eta$ is assumed. The optimal key rate scales quadratically with $\eta$. The corresponding value of the mean photon pair number parameter $\bar n$ depends on $\eta$ very weakly, allowing to choose one fixed value of $\bar n$ independently on $\eta$ to obtain key rate very close to the optimal value.}
	\label{fig:optima}
\end{figure}

The dependence of the optimal key rate on transmittance is shown in Fig.~\ref{fig:optima}. For easier depiction, we assume symmetric transmittances $\eta_\mathrm{A} = \eta_\mathrm{B} = \eta$. One can observe the dependence \eqref{eq:Rkeyopt}. The exact optimal point $\bar n$ depends on transmittance as well, albeit not significantly. This result means that setting the SPDC source at a certain gain is going to guarantee the optimal trade-off between entanglement and brightness.

\section{Experimental results}
To validate the predictions of our model, we used a CW pumped non-collinear, type-I SPDC \cite{Kwiat1999} with a BiBO nonlinear crystal. To obtain the maximal Bell factor without the necessity of measurement optimization we performed a full quantum state tomography on the source for varied lengths of coincidence windows, which allowed us to tune the mean pair number $\bar{n}$ and thus the rate $r_\mathrm{C}$ of the effective state $\rho$. The density matrices were each reconstructed from a set of 36 tomographic measurements using the method of maximum likelihood estimation \cite{Hradil1997, Jezek2003}. From these density matrices $r_\mathrm{DW}$ was calculated. Statistical confidence of each measurement was estimated by 2000 Monte-Carlo simulations based on Poissonian variance of all coincidence counts. The total number of coincidences $N_\mathrm{C}$ for coincidence rate calculation (\ref{eq:CW_coinc_rate}) was obtained by summing up coincidence counts for 4 complementary projections in each of the 9 tomographic sets of projections and then averaged, with $N_\mathrm{win} = T/\tau$ available from known duration of measurement $T$ and length of the coincidence window $\tau$. This is an accurate calculation of $N_\mathrm{C}$ for small multi-photon contributions, which holds for our data where $r_\mathrm{C} < 10^{-2}$. Both sets of $r_\mathrm{C}$ and $r_\mathrm{DW}$ allow us to compare the experimental data against the prediction of our model (see Fig.~\ref{fig:C_spdc_vs_dots_with_jwp}). For this prediction we choose the quantum state $\rho_0$ to be a density matrix of an entangled state produced by the source, which was obtained using a 1~ns coincidence window. This quantum state exhibits $S=2.815(5)$ and $Q=0.0013(5)$. Following the procedure outlined in Appendix~\ref{sec:CW_tomography_model} we arrive at different effective density matrices $\rho$ as the mean pair number $\bar{n}$ varies. The $\eta$ parameter of the model was set to the value of $0.16$ to reflect the two-photon collection efficiency of the experimental setup. A complete data set for the experimental points of Fig.~\ref{fig:C_spdc_vs_dots_with_jwp} is given in Table~\ref{tab:s_and_q_vs_tau} in Appendix~\ref{sec:data_table}.

\section{Comparison with quantum dot entanglement sources}
\begin{figure}[t]
	\centering\includegraphics[width=\columnwidth]{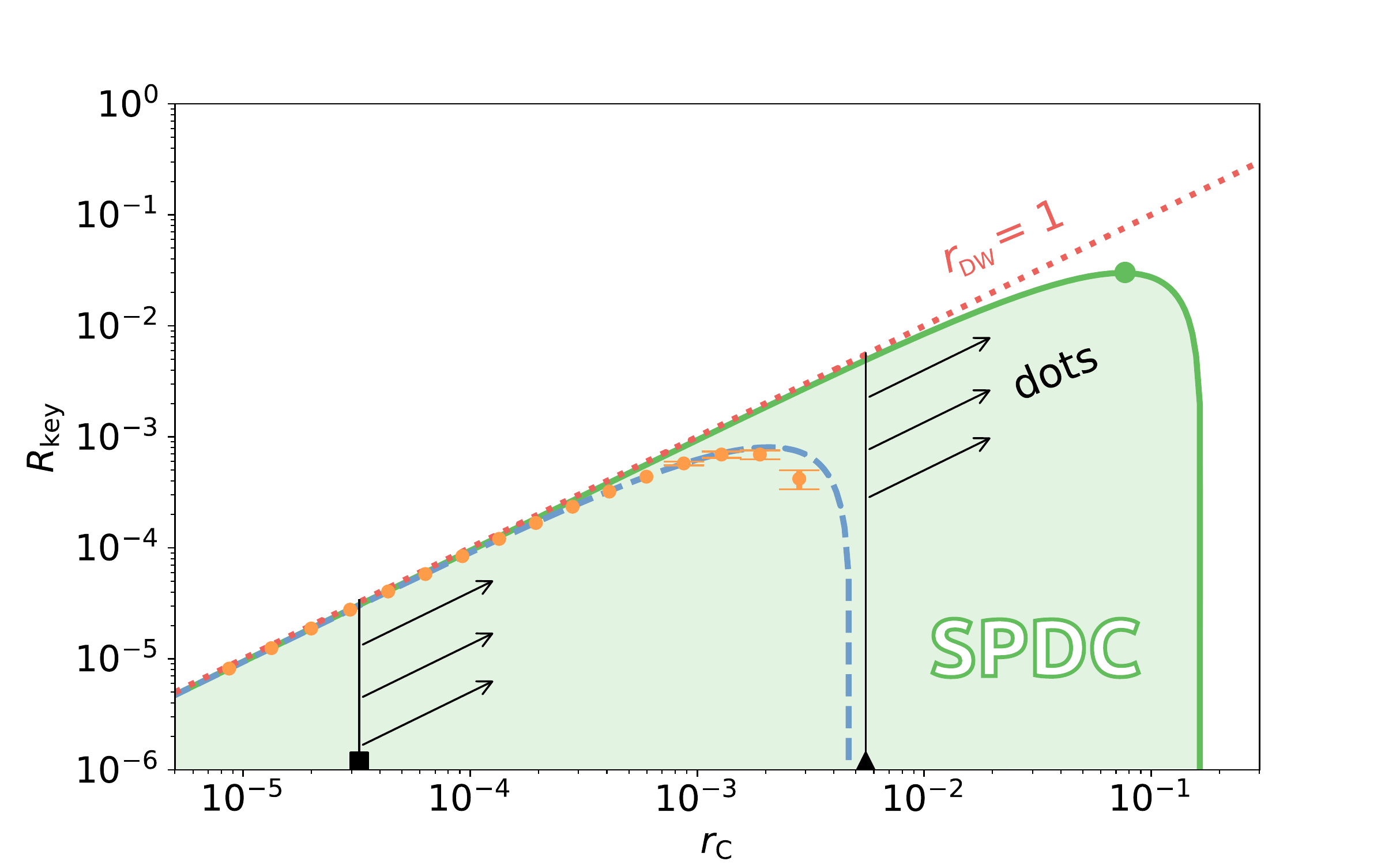}
	\caption{The dependence of key rate $R_\mathrm{key}$ on coincidence rate $r_\mathrm{C}$ for different implementations of quantum entanglement sources. The orange points represent density matrices reconstructed from a continuously pumped SPDC source, whereas the blue dashed line is the result of a CW SPDC model with $\eta_\mathrm{A}=\eta_\mathrm{B}=0.16$. The solid green line represents an ideal no-loss scenario and marks a fundamental limitation of SPDC entanglement sources in the QKD protocol. The green point shows the upper bound on key rate for SPDC entanglement sources, with $R_\mathrm{key}^\mathrm{max} = 0.029$. The black square and triangle marks represent the $r_\mathrm{C}$ of quantum dot entanglement sources \cite{Basset2019} and \cite{Wang2019}, respectively. The arrows pointing diagonally show that key rate of quantum dot entanglement sources increases linearly with $r_\mathrm{C}$. The red dotted line shows the linear dependence for an ideal quantum dot source with $r_\mathrm{DW}=1$. For such an ideal quantum dot source the dependence overlaps with that of an SPDC source for lower values of $r_\mathrm{C}$. For high enough $r_\mathrm{DW}$ and $r_\mathrm{C}$, quantum dot entanglement sources are going to surpass even the best SPDC sources.}
	\label{fig:C_spdc_vs_dots_with_jwp}
\end{figure}

Finally, SPDC is compared with recent entanglement sources based on quantum dots. Due to their discrete energy structure and strong sub-Poissonian nature of light emitted individually in the signal and idler mode, there is no multi-pair component in the generated entangled state. In addition, the generation of photon pairs can be achieved with near-unity efficiency. However, current sources often suffer from imperfect collection of photons, which in turn leads to increased losses and low coincidence rate. This means that improving collection efficiency is an important goal of quantum dot entanglement source engineering. The behavior of the key rate $R_\mathrm{key}$ with respect to $r_\mathrm{C}$ is shown in Fig.~\ref{fig:C_spdc_vs_dots_with_jwp}.

The $r_\mathrm{DW}$ of quantum dot entanglement sources is not subject to a fundamental quality-quantity trade-off, contrary to SPDC sources. The $r_\mathrm{DW}$ obtained using quantum dots depends primarily on achieved degree of entanglement and does not deteriorate with the increased excitation rate.
We illustrate this behavior in Fig.~\ref{fig:C_spdc_vs_dots_with_jwp}. The most noticeable feature is that the SPDC sources are systematically bounded whereas quantum dot ones are not. For a quantum dot source with known degree of entanglement \cite{Basset2019} the key rate will scale linearly with $r_\mathrm{C}$. With increased collection efficiency a quantum dot source can reach $r_\mathrm{C}$ that is above the one at which our experimental SPCD source can viably yield a non-zero $R_\mathrm{key}$.  Ongoing improvements in quantum dot entangled photon pair sources can be seen in their recent realizations \cite{Basset2019, Wang2019} where collection efficiency and quality of entanglement have been increased.

\section{Conclusion}
We predicted the dependence of key rate in entanglement-based QKD on the generation rate of photon-pair sources based on continuous-wave SPDC and enabled their comparison with quantum dot sources of entanglement. The SPDC is systematically limited by multiple photon pairs being generated during a single detection window. The multi-photon contribution corresponds to white noise proportional to the SPDC gain. Consequently, the secure key rate is fundamentally bound by the value $R_\mathrm{key}^\mathrm{max} = 0.029$ bits/window. The optimal gain for SPDC was shown to be $\bar{n}_\mathrm{opt} = 0.0737$ pairs/window.
 
Quantum dot sources, on the other hand, do not have a fundamental limit that is bound to photon statistics. This means that an increase of the coincidence rate $r_\mathrm{C}$ should not reduce the quality of the source, and with it related Devetak-Winter rate $r_\mathrm{DW}$. Therefore, it is reasonable to expect that quantum dot sources could overcome the SPDC ones in performance. How superior they can be depends on several factors, some of them limiting. Firstly, there is the source emission efficiency that is affected by phonons \cite{Denning2019}. These phonon effects can be minimized by embedding the quantum dot in a narrowband cavity. However, an efficient extraction of photon pairs demands broadband cavities, limiting the emission efficiency to 90~\%. There are various proposed schemes that target an ideal broadband cavity that allows for near unity collection of the emitted photons \cite{Osterkryger2019}, promising efficiencies of up to 99~\%. The state preparation could also have a limit in biexciton binding energy \cite{Huber2016} which could be overcome by adequate preparation of the excitation pulse. However, efficiency is not the only parameter determining the key rate $R_\mathrm{key}$. The currently achievable degree of entanglement is relatively high. Furthermore, the indistinguishability of the excitonic states would be further improved by embedding the quantum dot in a structure that features a high Purcell factor \cite{Osterkryger2019}. If we consider only dephasing noise and a concurrence of 95~\%, the minimum coincidence rate per excitation necessary to overcome the SPDC upper bound would be $r_\mathrm{C}>0.035$; $0.044$ in the case of a white noise. The secure key rate of quantum dot sources therefore has the potential to overcome SPDC following a number of technical optimizations.

The proposed quantification of a key rate per window could be extended to a key rate per time. This would include a multiplication by the number of detection windows per time, which is usually limited by the temporal resolution of the detectors. In the case of quantum dots, the lifetimes of the photons -- typically in the order of $\sim100$ ps -- represent the limit for the excitation frequency and for the coincidence rate per time. SPDC, on the other hand, can easily get the biphoton coherence to picosecond range, which is the current resolution limit of single-photon detectors \cite{Korzh2020Mar}. As a result, SPDC can benefit from narrower coincidence windows and therefore have an additional advantage in terms of key rate per time.

The authors have recently become aware of a recent work which demonstrates the feasibility of experimental employment of quantum dot entanglement sources for the purposes of QKD \cite{Basset2020}.

\begin{acknowledgments}
This work has received national funding from the MEYS and the funding from European Union's Horizon 2020 research and innovation framework programme under grant agreement No. 731473 (project 8C18002). Project Hyper-U-P-S has received funding from the QuantERA ERA-NET Cofund in Quantum Technologies implemented within the European Union's Horizon 2020 Programme. This work was also supported by the Czech Science Foundation (17-26143S). R.H. acknowledges the support of the Palacky University (project IGA-PrF-2020-009).
R.F. acknowledges the project CZ.02.1.01/0.0/0.0/16\_026/0008460 of MEYS CR. A.P. would also like to acknowledge Swedish Research Council and Carl Tryggers Stiftelse.
\end{acknowledgments}

\appendix

\section{Tomography in CW}
\label{sec:CW_tomography_model}

The effect of multi-photon contributions will be modeled in this section. The initial quantum state $\rho_0$ corresponds to a single photon pair and represents the low-gain-limit of the SPDC process. The state $\rho_0$ is subjected to a set of tomographic measurements. Each qubit is projected onto a state $\ket{\psi_i}$, typically one of the polarization states H, V, D, A, R, L. This would normally lead to a set of 36 two-qubit projections $c^0_{ij} = \bra{\psi_i \psi_j}\rho_0\ket{\psi_i \psi_j}$.

Here we also need the reduced one-qubit density matrices $\rho_0^\mathrm{A},\rho_0^\mathrm{B}$ by tracing over the other mode,
\begin{equation}
\rho_0^\mathrm{A/B} = \Tr_\mathrm{B/A}[\rho_0].
\end{equation}
For all projections $\{i,j\}$, we need to calculate probabilities of each detector clicking (1) or not clicking (0). Because the transmittances $\eta_\mathrm{A,B}$ may cause photons to be lost, the possible combinations are
\begin{align}
\label{eq:low_gain_pij_loss}
p_{ij}^{(11)} &= \eta_\mathrm{A} \eta_\mathrm{B} \bra{\psi_i \psi_j}\rho_0\ket{\psi_i \psi_j}, \\
p_{ij}^{(10)} &= \eta_\mathrm{A} \eta_\mathrm{B} \bra{\psi_i \psi_j^\perp}\rho_0\ket{\psi_i \psi_j^\perp}	\\\nonumber
 & + \eta_\mathrm{A}(1-\eta_\mathrm{B})\bra{\psi_i}\rho_0^\mathrm{A}\ket{\psi_i}, \\
p_{ij}^{(01)} &= \eta_\mathrm{A} \eta_\mathrm{B} \bra{\psi_i^\perp \psi_j}\rho_0\ket{\psi_i^\perp \psi_j}	\\\nonumber
 & + (1-\eta_\mathrm{A})\eta_\mathrm{B}\bra{\psi_j}\rho_0^\mathrm{B}\ket{\psi_j}, \\
p_{ij}^{(00)} &= \eta_\mathrm{A} \eta_\mathrm{B} \bra{\psi_i^\perp \psi_j^\perp}\rho_0\ket{\psi_i^\perp \psi_j^\perp}	\\\nonumber
 & + \eta_\mathrm{A}(1-\eta_\mathrm{B})\bra{\psi_i^\perp}\rho_0^\mathrm{A}\ket{\psi_i^\perp} \\\nonumber
 & + (1-\eta_\mathrm{A})\eta_\mathrm{B}\bra{\psi_j^\perp}\rho_0^\mathrm{B}\ket{\psi_j^\perp} 	\\\nonumber
 & + (1 - \eta_\mathrm{A}) (1 - \eta_\mathrm{B}),
\end{align}
with $\left\langle \psi_i | \psi_{i}^\perp \right\rangle = 0$. The order of the modes was set to $\rho_0 \in \mathcal{H}_\text{Alice} \otimes \mathcal{H}_\text{Bob}$, whereas the other variant can be expressed by swapping A and B.

In a real tomographic measurement, only coincidences (11) are registered, but they can be caused by multiple pairs. Assuming a short coherence time, the number of generated pairs $n$ follows the Poisson distribution. Then, the probability of a coincidence is
\begin{align}
\label{eq:c_ij}
c_{ij} &= \sum_{n=0}^\infty \left[1 - \left(p_{ij}^{(10)} + p_{ij}^{(00)}\right)^n - \left(p_{ij}^{(01)} + p_{ij}^{(00)}\right)^n \right. \\\nonumber
 & + \left. \left( p_{ij}^{(00)} \right)^n \right] \times \frac{n^{\bar{n}}}{n!}\mathrm{e}^{-\bar{n}}.
\end{align}
The mean pair number $\bar{n}$ is a parameter of the model proportional to the coincidence window width $\tau$ and the gain of the SPDC process. The density matrix $\rho$ is then found as the maximum-likelihood estimation that best explains the measured set of probabilities $\{c_{ij}\}$ \cite{Hradil1997, Jezek2003}.

If the initial state is chosen as one of the Bell states $\rho_0 = \rho_\mathcal{B}$, the result has the form
\begin{equation}
\label{eq:rho_kappa}
\rho = \left(1-\kappa\right) \rho_\mathcal{B} + \kappa \frac{1}{4} \mathds{1}\otimes\mathds{1}, \quad \kappa \in [0,1].
\end{equation}
The tomography of the state $\rho$ yields model probabilities that can be analytically calculated,
\begin{equation}
C_{ij} = \bra{\psi_i \psi_j}\rho\ket{\psi_i \psi_j}.
\end{equation}
The log-likelihood function then is (see Appendix~\ref{sec:likelihood})
\begin{equation}
\log\mathcal{L} = \sum_{i,j} c_{ij} \log(C_{ij}).
\end{equation}
The parameter $\kappa$ is obtained by solving $\partial(\log\mathcal{L})/\partial\kappa = 0$,
\begin{equation}
\label{eq:kappa}
\kappa = \frac{2\left(\mathrm{e}^\frac{\eta_\mathrm{A}\bar{n}}{2} - 1\right)\left(\mathrm{e}^\frac{\eta_\mathrm{B}\bar{n}}{2} - 1\right)}{1 - 2\,\mathrm{e}^\frac{\eta_\mathrm{A}\bar{n}}{2} - 2\,\mathrm{e}^\frac{\eta_\mathrm{B}\bar{n}}{2} + \mathrm{e}^\frac{\eta_\mathrm{A}\eta_\mathrm{B}\bar{n}}{2} + 2\,\mathrm{e}^\frac{\left(\eta_\mathrm{A} + \eta_\mathrm{B}\right)\bar{n}}{2}}.
\end{equation}

From (\ref{eq:rho_kappa}) it is possible to arrive to analytical expressions for the CHSH polynomial $S$ and QBER $Q$, which in turn can be used to calculate $r_\mathrm{DW}$:
\begin{eqnarray}
\label{eq:S_and_Q_and_rDW_kappa}
S &=& 2 \sqrt{2\left(1 - \kappa\right)^2},\quad Q = \frac{\kappa}{2}, \\
r_\mathrm{DW} &=& 1 - h(Q) - h\left(\frac{1 + \sqrt{\left(S/2\right)^2 - 1}}{2}\right).
\end{eqnarray}
The coincidence rate $r_\mathrm{C}$ is
\begin{align}\nonumber
r_\mathrm{C} &= \sum_{n=0}^\infty \left( 1 - \left( 1 - \eta_\mathrm{A} \right)^n \right) \left( 1 - \left( 1 - \eta_\mathrm{B} \right)^n \right) \times  \frac{n^{\bar{n}}}{n!}\mathrm{e}^{-\bar{n}} 	\\\label{eq:rc_kappa}
 & = 1 - \mathrm{e}^{-\eta_\mathrm{A}\bar{n}} - \mathrm{e}^{-\eta_\mathrm{B}\bar{n}} + \mathrm{e}^{-(\eta_\mathrm{A} + \eta_\mathrm{B} - \eta_\mathrm{A} \eta_\mathrm{B}) \bar{n}}.
\end{align}

\section{Likelihood definition}
\label{sec:likelihood}

To recapitulate, the objective of the quantum state tomography is to find a suitable density matrix $\rho$ that best explains the relative frequencies (probabilities) $\{c_{ij}\}$ that were either measured or modeled by \eqref{eq:c_ij}. Likelihood is a probability of obtaining the results $\{c_{ij}\}$ conditional on $\rho$. We denote the density matrix $\rho$ that maximizes the probability of obtaining the relative frequencies $\{c_{ij}\}$ as the maximum-likelihood estimate.

Let us suppose that we are running $N$ two-qubit projection measurements in a two-qubit basis $\{ |\varPsi_k\rangle \}_{k=1}^4$, where $\sum_k |\varPsi_k\rangle\langle\varPsi_k| = \mathds{1} \otimes \mathds{1}$. Then we obtain the projection probabilities
\begin{equation}
C_k = \langle \varPsi_k| \rho | \varPsi_k \rangle,
\end{equation}
where $\sum_k C_k = 1$. Such a measurement run would result in $n_k$ projections of each $|\varPsi_k\rangle$, ($\sum_k n_k = N$), and the probability of this result follows the multinomial distribution,
\begin{equation}
\label{eq:Pr_nk}
\Pr[\{n_k\}] = \frac{N!}{\prod_k (n_k!)} \prod_k C_k^{n_k}.
\end{equation}

In our model, we consider relative frequencies $c_k$ rather than counts $n_k$, which would be obtained for $N \to \infty$, and we denote
\begin{equation}
c_k \simeq \frac{n_k}{N}.
\end{equation}

The likelihood of obtaining $\{c_k\}$ when measuring in basis $\{|\varPsi_k\rangle\}$ then follows from \eqref{eq:Pr_nk},
\begin{equation}\label{eq.likelihood_psi}
\mathcal{L}_{\varPsi} = \frac{N!}{\prod_k (N c_k)!} \prod_k C_k^{N c_k}.
\end{equation}

The quantum state tomography consists of multiple projection bases, most commonly nine that correspond to all possible products of Pauli matrices. So, the overall probability (likelihood) of obtaining $\{c_{ij}\}$ is
\begin{equation}\label{eq.likelihood}
\mathcal{L} = \prod_\varPsi \mathcal{L}_\varPsi,
\end{equation}
where indexing over $\varPsi$ and $k$ just becomes indexing over $i,j$ in the manuscript.

To maximize the likelihood $\mathcal{L}$, it is more convenient to maximize $\log(\mathcal{L})$. These are equivalent, because logarithm is a monotonously increasing function. This lets us rewrite \eqref{eq.likelihood_psi} and \eqref{eq.likelihood} as 
\begin{equation}\label{eq.log_lik}
\log{\mathcal{L}} = N\sum_{ij} c_{ij} \log C_{ij} + \sum_\varPsi \log(N!) - \sum_{ij} \log[(N c_{ij})!].
\end{equation}
Now we find the maximum-likelihood estimate $\rho$ (represented by $C_{ij}$) by solving for the parameter $\kappa$,
\begin{equation}
\frac{\mathrm{d}\log\mathcal{L}}{\mathrm{d}\kappa} = 0.
\end{equation}
For the maximization, the likelihood does not have to be normalized (we leave out the factor $N$), and we can also omit constant terms that do not depend on $C_{ij}$ (the second and third sums in \eqref{eq.log_lik}). The likelihood is then simplified to the common form \cite{Jezek2003}
\begin{equation}
\log\mathcal{L} = \sum_{ij} c_{ij} \log C_{ij}.
\end{equation}

\section{Experimental cookbook}
\label{sec:cookbook}

Following the approach introduced in Ref. \cite{Horodecki1995}, let us formulate the optimal configuration of Alice's and Bob's bases $\rm A_0, A_1, A_2, B_1, B_2$ \cite{Pironio2009} given a reconstructed quantum state $\rho$. The bases $\rm A_1, A_2, B_1, B_2$ need to give the maximum CHSH violation \cite{Kofman2012Feb} and the bases $\rm A_0, B_1$ need to minimize the QBER. The respective derivations are presented in Appendix~\ref{sec:bases}.

We denote the measurement in basis $X$ by the operator $X = \ketbra{\psi}{\psi} - \ketbra{\psi^\perp}{\psi^\perp}$, where the direction of $|\psi\rangle$ can be parameterized using the unit vector $\boldsymbol{x} \in \mathbb{R}^3$ and the vector of Pauli matrices $\boldsymbol{\sigma} = \{\sigma_1, \sigma_2, \sigma_3\}$: $X = \boldsymbol{x} \cdot \boldsymbol{\sigma}.$ The bases will therefore be given by real unit vectors $\boldsymbol{a}_{0,1,2}$ for Alice and $\boldsymbol{b}_{1,2}$ for Bob.

We begin by introducing the real tensor $T_\rho$ and the positive symmetric tensor $U_\rho$ \cite{Horodecki1995}:
\begin{align}
T_{\rho,ij} &= \Tr[\rho \cdot (\sigma_i \otimes \sigma_j)],	\\
U_\rho &= T_\rho^\text{T}T_\rho.
\end{align}
Let us find the eigenvalues and unit eigenvectors of $U_\rho$,
\begin{equation}
U_\rho \boldsymbol{e}_i = \lambda_i \boldsymbol{e}_i, \quad |\boldsymbol{e}_i| = 1, \quad i=1,2,3,
\end{equation}
and index them in the descending order $\lambda_1 \geq \lambda_2 \geq \lambda_3$.

\newlength{\rsideW}
\settowidth{\rsideW}{$\displaystyle \sqrt{\frac{\lambda_1}{\lambda_1 + \lambda_2}}\frac{T_\rho \boldsymbol{e}_1}{\left| T_\rho \boldsymbol{e}_1 \right|} +  \sqrt{\frac{\lambda_2}{\lambda_1 + \lambda_2}}\frac{T_\rho \boldsymbol{e}_2}{\left| T_\rho \boldsymbol{e}_2 \right|}$}

The optimal choice of bases depends on which modes are assigned to Alice and Bob (see Appendix~\ref{sec:bases}).
\begin{enumerate}
\item $\rho \in \mathcal{H}_\text{Alice} \otimes \mathcal{H}_\text{Bob}$
\begin{align}
\label{CHSHbasis.first}
\boldsymbol{a}_0 &= \frac{T_\rho \boldsymbol{e}_1}{\left| T_\rho \boldsymbol{e}_1 \right|}	\\
\label{CHSHbasis.2}
\boldsymbol{a}_1 &= \sqrt{\frac{\lambda_1}{\lambda_1 + \lambda_2}}\frac{T_\rho \boldsymbol{e}_1}{\left| T_\rho \boldsymbol{e}_1 \right|} +  \sqrt{\frac{\lambda_2}{\lambda_1 + \lambda_2}}\frac{T_\rho \boldsymbol{e}_2}{\left| T_\rho \boldsymbol{e}_2 \right|}	\\
\label{CHSHbasis.3}
\boldsymbol{a}_2 &= \sqrt{\frac{\lambda_1}{\lambda_1 + \lambda_2}}\frac{T_\rho \boldsymbol{e}_1}{\left| T_\rho \boldsymbol{e}_1 \right|} -  \sqrt{\frac{\lambda_2}{\lambda_1 + \lambda_2}}\frac{T_\rho \boldsymbol{e}_2}{\left| T_\rho \boldsymbol{e}_2 \right|}	\\
\label{CHSHbasis.4}
\boldsymbol{b}_{1,2} &= \boldsymbol{e}_{1,2}
\end{align}

\item $\rho \in \mathcal{H}_\text{Bob} \otimes \mathcal{H}_\text{Alice}$
\begin{align}
\label{CHSHbasis.5}
\boldsymbol{a}_0 &= \makebox[\rsideW][l]{$\boldsymbol{e}_1$}	\\
\label{CHSHbasis.6}
\boldsymbol{a}_1 &= \sqrt{\frac{\lambda_1}{\lambda_1 + \lambda_2}}\boldsymbol{e}_1 +  \sqrt{\frac{\lambda_2}{\lambda_1 + \lambda_2}}\boldsymbol{e}_2	\\
\label{CHSHbasis.7}
\boldsymbol{a}_2 &= \sqrt{\frac{\lambda_1}{\lambda_1 + \lambda_2}}\boldsymbol{e}_1 -  \sqrt{\frac{\lambda_2}{\lambda_1 + \lambda_2}}\boldsymbol{e}_2	\\
\label{CHSHbasis.last}
\boldsymbol{b}_{1,2} &= \frac{T_\rho \boldsymbol{e}_{1,2}}{\left| T_\rho \boldsymbol{e}_{1,2} \right|} 
\end{align}

\end{enumerate}

The optimal quantities are given by the two largest eigenvalues,
\begin{align}\label{CHSH.max}
S &= 2\sqrt{\lambda_1 + \lambda_2},	\\
Q &= \frac{1 - \sqrt{\lambda_1}}{2}.
\end{align}

\begin{figure}
\centering
\vspace{5mm}
\includegraphics[width=\columnwidth]{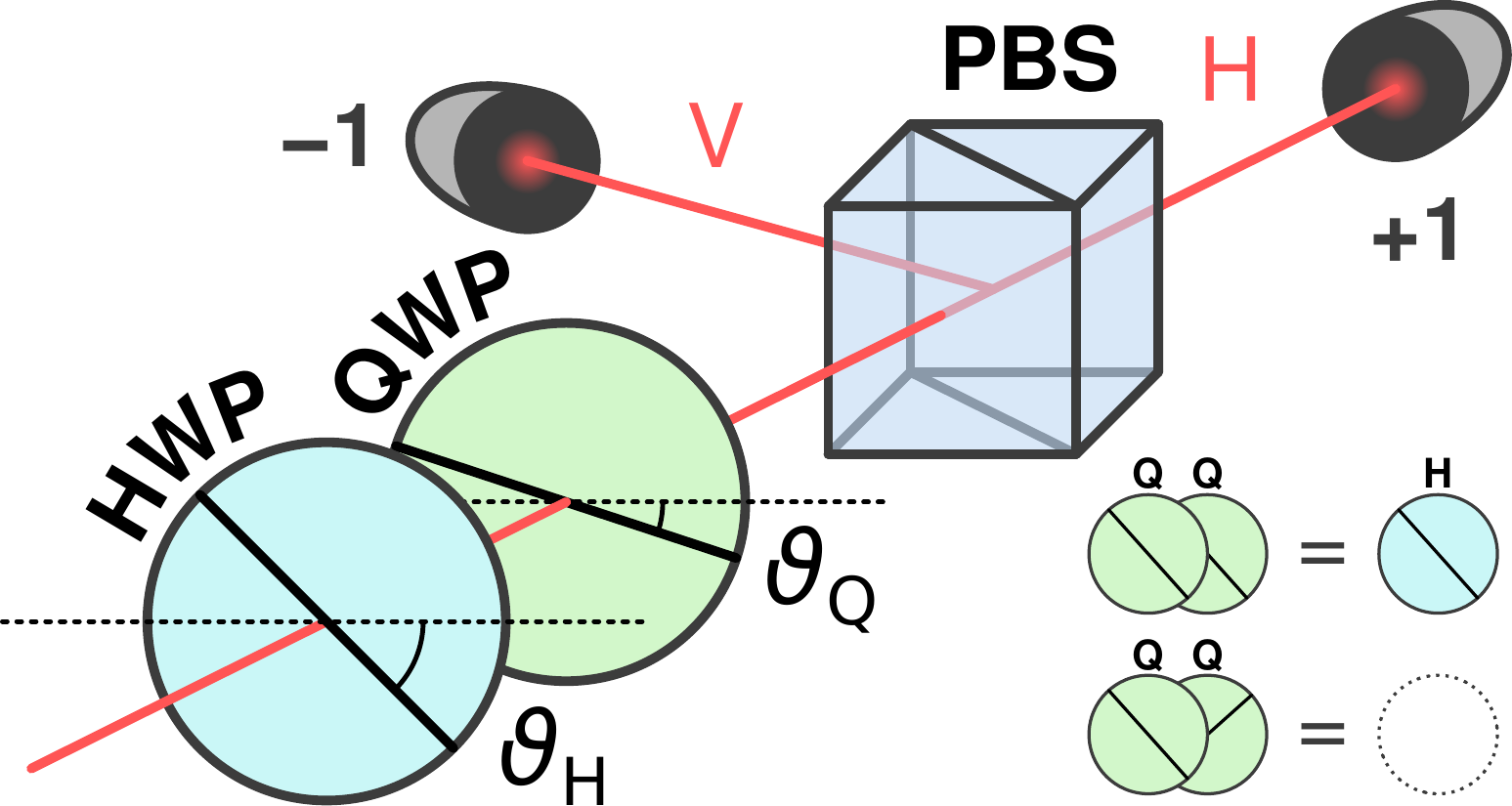}
\caption{Polarization projection measurement using two waveplates (half-wave -- HWP, and quarter-wave -- QWP) and a polarizing beam splitter (PBS). Let us assign the outcome $+1$ to a detection in the horizontal output and $-1$ to the the vertical output. The angles $\vartheta_\text{H}, \vartheta_\text{Q}$ are between the respective waveplate axes and the horizontal plane. The HWP and QWP angles exhibit periodicity: $\vartheta_\text{H} \Leftrightarrow \vartheta_\text{H} + k \pi/2, \vartheta_\text{Q} \Leftrightarrow \vartheta_\text{Q} + l \pi; k,l \in \mathbb{Z}$.}
\label{fig:projectionwaveplates}
\end{figure}

Upon obtaining a basis vector $\boldsymbol{x} = \{ x_1, x_2, x_3 \}$, an experimentalist needs to know how to set up the polarization measurement. Let us suppose that our basis is chosen in a horizontal/vertical polarization so that $\sigma_1 = \sigma_x$, $\sigma_2 = \sigma_y$, $\sigma_3 = \sigma_z = \ketbra{\text{H}}{\text{H}} - \ketbra{\text{V}}{\text{V}}$. Also, let us assume the projection set-up shown in Fig.~\ref{fig:projectionwaveplates}.

Then, the waveplate axes rotations with respect to the horizontal plane can be obtained by
\begin{align}
\vartheta_\text{Q} &= \frac{1}{2}\arcsin(x_2),	\\
\vartheta_\text{H} &= \frac{1}{4}\left[\arctan\left(\frac{x_1}{x_3}\right)+\arcsin(x_2)\right].
\end{align}

There is an important caveat about quarter-wave plates of Alice and Bob. While the choice of slow or fast axis is arbitrary for all waveplates, the QWP axes need to be oriented consistently in both modes. That is, for $\vartheta_\text{Q} = 0$, both Alice's and Bob's QWPs need to have either both their fast axes horizontal, or both their slow axes horizontal. Since it is easy to calibrate the directions of all axes up to $\pi/2$ using linear polarizers, the two QWPs only need to be matched together. This can be achieved by aligning any of their axes and rotating both QWPs simultaneously. If their slow and fast axes are parallel, the overall transformation corresponds to a HWP. If the axes are perpendicular, the waveplates cancel each other out and no polarization modulation occurs (Fig.~\ref{fig:projectionwaveplates}).

\section{Alice's and Bob's optimal choice of bases}
\label{sec:bases}

Eqs. \eqref{CHSHbasis.first} to \eqref{CHSHbasis.last} are the result of a conjunction of two conditions -- minimizing $Q$ and maximizing $S$. Here we present the respective derivations.

\subsection*{Optimal QBER}

The optimal QBER is obtained by the same principle as described in Ref. \cite{Horodecki1995}. The definition follows from \eqref{eq:qber}, assuming projection measurements in bases $\mathrm{Q}_1, \mathrm{Q}_2$, and can be written as
\begin{align}
\label{eq:our_qber}
Q &= \mathrm{Tr}\left[\rho\left(\hat\Pi_{\mathrm{Q}_1}^{(+)}\otimes\hat\Pi_{\mathrm{Q}_2}^{(-)}\right)\right] \\\nonumber
& + \mathrm{Tr}\left[\rho\left(\hat\Pi_{\mathrm{Q}_1}^{(-)}\otimes\hat\Pi_{\mathrm{Q}_2}^{(+)}\right)\right],
\end{align}
where $\hat\Pi_{\mathrm{Q}_1}^{(\pm)}$ and $\hat\Pi_{\mathrm{Q}_2}^{(\pm)}$ are the projector operators onto the $(+)$ or $(-)$ states in the respective bases. Using the real-vector formalism, the bases are given by $\boldsymbol{q}_1, \boldsymbol{q}_2$, and the QBER operator is
\begin{align}
\mathcal{Q} &= \frac{1}{2}\left[ \mathds{1}\otimes\mathds{1} - (\boldsymbol{q}_1\cdot\boldsymbol{\sigma}) \otimes (\boldsymbol{q}_2\cdot\boldsymbol{\sigma}) \right],	\\
Q &= \Tr[\mathcal{Q}\rho].
\end{align}
If we rewrite the QBER as
\begin{equation}
Q = \frac{1}{2}(1-\boldsymbol{q}_1^\text{T}\cdot T_\rho \cdot \boldsymbol{q}_2),
\end{equation}
the optimum requires maximizing the second term. The inner product of $\boldsymbol{q}_1$ and $T_\rho \boldsymbol{q}_2$ is clearly maximized by choosing the unit vector $\boldsymbol{q}_1$ to be in the same direction, $\boldsymbol{q}_1 = T_\rho \boldsymbol{q}_2/|T_\rho \boldsymbol{q}_2|$. If follows that
\begin{equation}
Q = \frac{1}{2}\left(1- \sqrt{\boldsymbol{q}_2^\text{T}U_\rho \boldsymbol{q}_2}\right).
\end{equation}
By considering $\boldsymbol{q}_2$ in the eigenbasis of $U_\rho$, the maximum of the product can be easily found to be the largest eigenvalue of $U_\rho$, that is $\lambda_1$, and so $\boldsymbol{q}_2 = \boldsymbol{e}_1$. If there are multiple maximum eigenvalues, such as for the Bell states, the vector $\boldsymbol{q}_2$ may belong to any subspace spanned by the corresponding eigenvectors. This result corresponds to the equations \eqref{CHSHbasis.first}, \eqref{CHSHbasis.4}, \eqref{CHSHbasis.5}, and \eqref{CHSHbasis.last}.

\subsection*{Optimal CHSH violation}

The derivation follows the $U_\rho$-matrix approach outlined in \cite{Horodecki1995}, and was explicitly calculated in \cite{Kofman2012Feb}. Let us follow the procedure in \cite{Horodecki1995} adopting the notation of $\boldsymbol{a}$ and $\boldsymbol{b}$ corresponding to $\rho \in \mathcal{H}_\text{Alice} \otimes \mathcal{H}_\text{Bob}$. Following the definition of the CHSH polynomial given for the protocol in Ref.~\cite{Pironio2009}, we obtain
\begin{equation}
\label{eq:S}
S = \boldsymbol{a}_1^\text{T} T_\rho (\boldsymbol{b}_1 + \boldsymbol{b}_2) + \boldsymbol{a}_2^\text{T} T_\rho (\boldsymbol{b}_1 - \boldsymbol{b}_2).
\end{equation}
We introduce two orthogonal unit vectors
\begin{align}\label{eq.c1}
\boldsymbol{c}_1 &= \frac{\boldsymbol{b}_1 + \boldsymbol{b}_2}{2\cos\theta},	\\
\label{eq.c2}
\boldsymbol{c}_2 &= \frac{\boldsymbol{b}_1 - \boldsymbol{b}_2}{2\sin\theta},
\end{align}
where $\theta \in (0,\pi/2)$ is half the angle between $\boldsymbol{b}_1$ and $\boldsymbol{b}_2$. This is just a different parameterization of $\boldsymbol{b}_1$ and $\boldsymbol{b}_2$, as any pair of such vectors can be represented by a unique combination of $\boldsymbol{c}_1$, $\boldsymbol{c}_2$, and the angle $\theta$; and vice versa. This allows us to rewrite the Bell factor as a sum of two scalar products,
\begin{equation}
S = [\boldsymbol{a}_1 \cdot (T_\rho \boldsymbol{c}_1)] 2\cos\theta + [\boldsymbol{a}_2 \cdot (T_\rho \boldsymbol{c}_2)] 2\sin\theta.
\end{equation}
Trivially, the maximum of each scalar product is reached for the unit vectors $\boldsymbol{a}_{1,2}$ that have the same direction as the right side,
\begin{equation}\label{eq.ai}
\boldsymbol{a}_{1,2} = \frac{T_\rho \boldsymbol{c}_{1,2}}{|T_\rho \boldsymbol{c}_{1,2}|}.
\end{equation}
This gives us
\begin{equation}
S = |T_\rho \boldsymbol{c}_1| 2\cos\theta + |T_\rho \boldsymbol{c}_2| 2\sin\theta.
\end{equation}
Now we maximize with respect to $\theta$ by solving $\partial S / \partial \theta = 0$, yielding
\begin{align}
S &= 2\sqrt{|T_\rho \boldsymbol{c}_1|^2 + |T_\rho \boldsymbol{c}_2|^2}, \\
\label{eq.theta}
\tan \theta &= \frac{|T_\rho \boldsymbol{c}_2|}{|T_\rho \boldsymbol{c}_1|}.
\end{align}
By denoting $U_\rho = T_\rho^\mathrm{T} T_\rho$, we can rewrite the norms and scalar products as
\begin{equation}\label{eq.Bell_cUc}
S = 2\sqrt{\boldsymbol{c}_1^\mathrm{T} U_\rho \boldsymbol{c}_1 + \boldsymbol{c}_2^\mathrm{T} U_\rho \boldsymbol{c}_2}.
\end{equation}

$U_\rho$ is a symmetrical non-negative diagonalizable matrix. The property of these matrices is that the sum of the products of two orthogonal unit vectors---such as in \eqref{eq.Bell_cUc}---is constant for all such vectors within a single plane. This is also related to invariance of the matrix trace under rotation around a coordinate axis.

The sum in \eqref{eq.Bell_cUc} can be maximized either using Lagrange multipliers and $\boldsymbol{c}_{1,2}$ taken in the eigenbasis of $U_\rho$, or using standard differential maximization of a rotated matrix $U_\rho$ in spherical coordinates and vectors $\boldsymbol{c}_{1,2}$ rotating in the $x$-$y$ plane.

The result is that maximal CHSH violation is reached for all orthogonal pairs of $\boldsymbol{c}_1$ and $\boldsymbol{c}_2$ in the plane corresponding to two greatest eigennumbers of $U_\rho$. If we maintain the notation of $\boldsymbol{e}_i$ being the eigenvectors of $U_\rho$ with the corresponding eigennumbers $\lambda_1 \geq \lambda_2 \geq \lambda_3$, then the solution is given by
\begin{align}\label{eq.c1_sol}
\boldsymbol{c}_1 &= \boldsymbol{e}_1 \cos\varphi + \boldsymbol{e}_2 \sin\varphi, \\
\label{eq.c2_sol}
\boldsymbol{c}_2 &= \boldsymbol{e}_1 \sin\varphi - \boldsymbol{e}_2 \cos\varphi,
\end{align}
where $\varphi$ is an arbitrary angle \cite{Kofman2012Feb}. The bases $\boldsymbol{a}_{1,2}, \boldsymbol{b}_{1,2}$ are then obtained using equations \eqref{eq.c1}, \eqref{eq.c2}, \eqref{eq.ai}, and \eqref{eq.theta}.

The optimal bases have at least one degree of freedom ($\varphi$; more degrees if any two eigennumbers $\lambda_i$ are equal). Alice's and Bob's projections belong to two respective planes on the Bloch sphere that are generally different. $\boldsymbol{b}_{1,2}$ belong to the plane spanned by $\boldsymbol{c}_{1,2}$ due to the above-mentioned rotational symmetry of $\boldsymbol{c}_{1,2}$ under the free parameter $\varphi$. The $T_\rho$ image of this plane contains all vectors $\boldsymbol{a}_{1,2}$, owing to the property of linear transformations mapping planes unto planes.

\subsection*{Conjunction for \boldmath{$\rho \in \mathcal{H}_\text{Alice} \otimes \mathcal{H}_\text{Bob}$}}

As we showed above, the optimal QBER requires
\begin{align}
\boldsymbol{a}_0 &= \boldsymbol{q}_1 =  T_\rho \boldsymbol{e}_1/|T_\rho \boldsymbol{e}_1|, \\
\label{eq:cond.b1}
\boldsymbol{b}_1 &= \boldsymbol{q}_2 = \boldsymbol{e}_1.
\end{align}
From Eqs.~\eqref{eq.c1}, \eqref{eq.c2}, \eqref{eq.c1_sol}, and \eqref{eq.c2_sol}, we obtain
\begin{align}
\label{eq:b1}
\boldsymbol{b}_1 &= \cos(\varphi-\theta)\boldsymbol{e}_1 + \sin(\varphi-\theta)\boldsymbol{e}_2, \\
\label{eq:b2}
\boldsymbol{b}_2 &= \cos(\varphi+\theta)\boldsymbol{e}_1 + \sin(\varphi+\theta)\boldsymbol{e}_2.
\end{align}
Eq.~\eqref{eq:cond.b1} introduces a binding condition $\varphi = \theta$. To find $\theta$, we take \eqref{eq.theta} and remember that
\begin{equation}
|T_\rho\boldsymbol{c}_{1,2}| = \sqrt{\boldsymbol{c}_{1,2}^\mathrm{T}U_\rho\boldsymbol{c}_{1,2}}.
\end{equation}
When substituting \eqref{eq.c1_sol} and \eqref{eq.c2_sol}, we arrive at
\begin{equation}
\tan\theta = \sqrt{\frac{\lambda_1 \tan\theta + \lambda_2}{\lambda_1 + \lambda_2 \tan\theta}}.
\end{equation}
Since $\theta \in (0,\pi/2)$, the result leads to a quadratic equation with a single solution
\begin{equation}
\varphi = \theta = \frac{\pi}{4}.
\end{equation}
Substituting into \eqref{eq:b1} and \eqref{eq:b2}, we obtain \eqref{CHSHbasis.4}. Eq.~\eqref{eq.ai} yields
\begin{equation}
\boldsymbol{a}_{1,2} = \frac{T_\rho(\boldsymbol{e}_1 \pm \boldsymbol{e}_2)}{\sqrt{\lambda_1 + \lambda_2}},
\end{equation}
which, after normalizing the vectors, leads to \eqref{CHSHbasis.2} and \eqref{CHSHbasis.3}, completing the optimal QKD bases.

\subsection*{Conjunction for \boldmath{$\rho \in \mathcal{H}_\text{Bob} \otimes \mathcal{H}_\text{Alice}$}}

In order to maintain mode consistency with the derivation of optimal CHSH and with the polynomial \eqref{eq:S}, we assign vectors $\widetilde{\boldsymbol{a}}$ to Bob and $\widetilde{\boldsymbol{b}}$ to Alice, with the tilde serving as a reminder to swap the notation eventually.

The QBER optimization reads
\begin{align}
\widetilde{\boldsymbol{a}}_1 &= \boldsymbol{q}_1 =  T_\rho \boldsymbol{e}_1/|T_\rho \boldsymbol{e}_1|, \\
\widetilde{\boldsymbol{b}}_0 &= \boldsymbol{q}_2 = \boldsymbol{e}_1.
\end{align}
Here, Bob's first basis $\widetilde{\boldsymbol{a}}_1$ introduces a binding condition, which, after looking at \eqref{eq.ai}, \eqref{eq.c1_sol}, and \eqref{eq.c2_sol}, simply gives
\begin{align}
\boldsymbol{c}_1 &= \boldsymbol{e}_1, \\
\boldsymbol{c}_2 &= \boldsymbol{e}_2, \\
\widetilde{\boldsymbol{a}}_2 &= T_\rho \boldsymbol{e}_2/|T_\rho \boldsymbol{e}_2|.
\end{align}

Like before, we substitute the vectors $\boldsymbol{c}_{1,2}$ into \eqref{eq.theta}, so we can have both angles solved,
\begin{align}
\varphi &= 0, \\
\label{eq:thetacond}
\theta &= \arctan\sqrt{\lambda_2/\lambda_1}.
\end{align}
For $\theta \in (0,\pi/2)$, we know that
\begin{align}
\label{eq:costheta}
\cos\theta &= \frac{1}{\sqrt{1+\tan^2\theta}}, \\
\label{eq:sintheta}
\sin\theta &= \frac{\tan\theta}{\sqrt{1+\tan^2\theta}}.
\end{align}
Substituting \eqref{eq:thetacond}, \eqref{eq:costheta}, and \eqref{eq:sintheta} into \eqref{eq.c1} and \eqref{eq.c2}, we straightforwardly solve for
\begin{equation}
\widetilde{\boldsymbol{b}}_{1,2} = \sqrt{\frac{\lambda_1}{\lambda_1+\lambda_2}}\boldsymbol{e}_1 \pm \sqrt{\frac{\lambda_2}{\lambda_1+\lambda_2}}\boldsymbol{e}_2.
\end{equation}

Now, by swapping the notation, $\widetilde{\boldsymbol{a}} \to \boldsymbol{b}$, $\widetilde{\boldsymbol{b}} \to \boldsymbol{a}$, we have derived the results \eqref{CHSHbasis.5} to \eqref{CHSHbasis.last}.

\section{Experimental data}
\label{sec:data_table}

\begin{table*}[ht]
%\begin{ruledtabular}
\setlength{\tabcolsep}{10pt}
	\begin{tabular}{r@{.}l l@{$\times$}l r@{.}l r@{.}l r@{.}l l@{$\times$}l}
		\hline
		\hline
		\multicolumn{2}{c}{$\tau$ [ns]} & \multicolumn{2}{c}{$r_\mathrm{C}$} & \multicolumn{2}{c}{$S$} & \multicolumn{2}{c}{$Q$}  &
		\multicolumn{2}{c}{$r_\mathrm{DW}$} & \multicolumn{2}{c}{$R_\mathrm{key}$}\\
		\hline
		1&0   & 8.66(3)&$10^{-6}$ & 2&815(5) & 0&0013(5) & 0&94(2) & 8.2(2)&$10^{-6}$ \\
		1&4   & 1.330(5)&$10^{-5}$ & 2&814(5) & 0&0015(6) & 0&94(2) & 1.25(2)&$10^{-5}$ \\
		2&1   & 1.992(7)&$10^{-5}$ & 2&814(5) & 0&0013(6) & 0&94(2) & 1.88(3)&$10^{-5}$ \\
		3&0   & 2.96(1)&$10^{-5}$  & 2&814(4) & 0&0016(6) & 0&94(2) & 2.77(5)&$10^{-5}$ \\
		4&3   & 4.35(1)&$10^{-5}$  & 2&812(5) & 0&0017(6) & 0&93(2) & 4.04(7)&$10^{-5}$ \\
		6&2   & 6.35(2)&$10^{-5}$  & 2&808(5) & 0&0022(8) & 0&92(2) & 5.8(1)&$10^{-5}$ \\
		8&9   & 9.23(3)&$10^{-5}$  & 2&807(5) & 0&0023(9) & 0&91(2) & 8.4(2)&$10^{-5}$ \\
		12&7  & 1.341(4)&$10^{-4}$  & 2&802(6) & 0&003(1)  & 0&90(2) & 1.20(3)&$10^{-4}$ \\
		18&3  & 1.946(7)&$10^{-4}$  & 2&790(6) & 0&004(1)  & 0&86(2) & 1.68(5)&$10^{-4}$ \\
		26&4  & 2.824(9)&$10^{-4}$  & 2&779(7) & 0&004(2)  & 0&83(2) & 2.35(7)&$10^{-4}$ \\
		37&9  & 4.10(1)&$10^{-4}$   & 2&763(7) & 0&007(2)  & 0&78(2) & 3.2(1)&$10^{-4}$ \\
		54&6  & 5.97(2)&$10^{-4}$   & 2&743(7) & 0&009(2)  & 0&73(2) & 4.4(1)&$10^{-4}$ \\
		78&5  & 8.72(3)&$10^{-4}$   & 2&716(8) & 0&012(3)  & 0&66(3) & 5.8(2)&$10^{-4}$ \\
		112&9 & 1.279(4)&$10^{-3}$   & 2&67(1)  & 0&020(4)  & 0&54(4) & 6.9(5)&$10^{-4}$ \\
		162&4 & 1.888(6)&$10^{-3}$   & 2&60(1)  & 0&033(4)  & 0&37(3) & 6.9(6)&$10^{-4}$ \\
		233&6 & 2.814(8)&$10^{-3}$   & 2&49(1)  & 0&053(4)  & 0&15(3) & 4.2(8)&$10^{-4}$ \\
		336&0 & 4.25(1)&$10^{-3}$    & 2&35(1)  & 0&080(4)  & \multicolumn{2}{c}{0}  & \multicolumn{2}{c}{0} \\
		483&3 & 6.53(2)&$10^{-3}$    & 2&18(1)  & 0&109(4)  & \multicolumn{2}{c}{0}  & \multicolumn{2}{c}{0} \\
		695&2 & 1.022(3)&$10^{-2}$    & 1&98(1)  & 0&146(3)  & \multicolumn{2}{c}{0}  & \multicolumn{2}{c}{0} \\
		1000&0& 1.641(4)&$10^{-2}$    & 1&73(1)  & 0&191(3)  & \multicolumn{2}{c}{0}  & \multicolumn{2}{c}{0} \\
		\hline
		\hline
	\end{tabular}
%\end{ruledtabular}
\caption{A list of $r_\mathrm{C}$, $S$, $Q$, $r_\mathrm{DW}$, and $R_\mathrm{key}$ values for the respective coincidence window lengths $\tau$ used during the experimental study of the dependence of $R_\mathrm{key}$ on $r_\mathrm{C}$.}
\label{tab:s_and_q_vs_tau}
\end{table*}

In the experiment, we used varying coincidence window lengths to study the effect of multi-photon components. In Table~\ref{tab:s_and_q_vs_tau}, we list the values of $S$, $Q$, $r_\mathrm{DW}$ and $R_{\text{key}}$ calculated from the tomographic reconstructions, for different values of the coincidence window length $\tau$. Experimentally obtained values of $r_\mathrm{C}$ are also included.

\end{document}